\documentclass[twocolumn,english,reprint, longbibliography, superscriptaddress, breaklinks=true, showkeys, showpacs=false, nofootinbib]{revtex4}
\usepackage[T1]{fontenc}
\usepackage[latin9]{inputenc}
\usepackage{color}
\usepackage{babel}
\usepackage{amsmath}
\usepackage{amsthm}
\usepackage{amsfonts}
\usepackage{amssymb}
\usepackage{graphicx}
\usepackage{subfigure}
\usepackage{physics}

\usepackage[unicode=true,pdfusetitle,
 bookmarks=true,bookmarksnumbered=false,bookmarksopen=false,
 breaklinks=true,pdfborder={0 0 0},backref=false,colorlinks=true]
 {hyperref} 
\usepackage{breakurl}

\makeatletter
\@ifundefined{textcolor}{}
{%
 \definecolor{BLACK}{gray}{0}
 \definecolor{WHITE}{gray}{1}
 \definecolor{RED}{rgb}{1,0,0}
 \definecolor{GREEN}{rgb}{0,1,0}
 \definecolor{BLUE}{rgb}{0,0,1}
 \definecolor{CYAN}{cmyk}{1,0,0,0}
 \definecolor{MAGENTA}{cmyk}{0,1,0,0}
 \definecolor{YELLOW}{cmyk}{0,0,1,0}
}

\pdfoutput=1
\hypersetup{colorlinks=true,citecolor=blue,linkcolor=cyan,urlcolor=blue,filecolor= green, breaklinks=true}
\usepackage{url}
\usepackage{breakurl}

\makeatother

\begin{document}

\title{The Sagnac effect for spin-$1/2$ particles through local Wigner rotations}

\author{Marcos L. W. Basso}
\email{marcoslwbasso@hotmail.com}
\thanks{corresponding author}
\address{Departamento de F\'isica, Centro de Ci\^encias Naturais e Exatas, Universidade Federal de Santa Maria, Avenida Roraima 1000, Santa Maria, Rio Grande do Sul, 97105-900, Brazil}
\address{New adress: Centro de Ci\^encias Naturais e Humanas, Universidade Federal do ABC, Avenida dos Estados 5001, 09210-580 Santo Andr\'e, S\~ao Paulo, Brazil}

\author{Jonas Maziero}
\email{jonas.maziero@ufsm.br}
\address{Departamento de F\'isica, Centro de Ci\^encias Naturais e Exatas, Universidade Federal de Santa Maria, Avenida Roraima 1000, Santa Maria, Rio Grande do Sul, 97105-900, Brazil}

\selectlanguage{english}%

\begin{abstract} 
 \textbf{Abstract:} In this article, we study the Sagnac effect for spin-$1/2$ particles through local Wigner rotations, according to the framework developed by [H. Terashima and M. Ueda, Phys. Rev. A 69, 032113 (2004)]. As the particles' spin works as a quantum `clock', when it moves in a superposition of co-rotating and counter-rotating circular paths in a rotating platform, its spin gets entangled with the momentum due to the local Wigner rotations. Therefore, in contrast with other works in the literature which showed that a rotating space-time (or a rotating frame) causes a shift in the interference pattern, here we show that a rotating spacetime also lead to a drop in the interferometric visibility, once there is a difference in the proper time elapsed along the two paths, which is known as the Sagnac time delay.
 
\end{abstract}


\maketitle

\section{Introduction}
Understanding the interconnections between gravity and quantum mechanics is a fascinating and difficult problem that physicists are facing nowadays. Notwithstanding, the lack of empirical evidence caused a discussion about whether gravity is a quantum entity or not. By its turn, this motivated a growing effort for probing the interplay between these two pillars of modern physics or, more precisely, to witness the quantumness of gravity \cite{Bose, Chiara, Carlo, Sougato, Vedral} and also for probing effects of general relativity in quantum phenomena \cite{Zych, Magdalena, Costa, Marcos}. For example,  the authors of Ref. \cite{Zych} regarded a Mach-Zehnder interferometer subjected to the gravitational potential of the Earth, where a `clock' is used as an interfering quanton.
Due to the difference in proper time elapsed along the two branches of the interferometer, the internal degree of freedom evolves to different quantum states for each path, what diminishes the interferometric visibility proportionally to the which-way information available in the final state of the clock, once the internal d.o.f gets entangled with the external d.o.f. of the particle. On the other side, in Ref. \cite{Vedral} the authors used a quantum variant of the Sagnac interferometer to argue for the quantum nature of gravity.

A long time ago, Sagnac predicted and experimentally verified that there exists a shift of the interference pattern when an interferometric apparatus is rotating, compared to what is observed when the device is at rest \cite{Sagnac, Rug}. Its applications are several, such as fiber optic gyroscopes, used in inertial navigation, and ring laser gyroscopes, used in geophysics \cite{Rugg}, as well in  the global positioning system \cite{Ashby}. Besides, there exists many experiments suggesting that the Sagnac effect is universal, in the sense that it is independent of the nature of the interfering beams \cite{Ml}. For instance, the Sagnac effect with matter waves has been verified experimentally using Cooper pairs \cite{Zim}, neutrons \cite{Atw}, and electrons \cite{Hassel}. In a series of remarkable experiments, Werner et al. \cite{Werner} demonstrated the effect of the terrestrial rotation on the neutrons' phase.

More recent experiments include the Hong-Ou-Mandel (HOM) interference on a rotating platform \cite{Marko}, and it was shown that the rotational motion of the platform induces a relative delay in the photon arrival times at the exit beam splitter and that this delay is observed as a shift in the position of the HOM dip. Also, in Ref. \cite{Toros} the authors have  proposed a modified HOM interferometer where entanglement can be revealed or concealed depending on the rotational frequency. Specifically, the authors have shown that rotations together with an asymmetry of the experimental setup can strongly affect the bunching and anti-bunching behavior, and hence the manifestation of entanglement. Finally, in Ref. \cite{Armin}, by performing a neutron interferometric experiment, the authors observed phase shifts arising as a consequence of the spin's coupling with the angular velocity of a rotating magnetic field. This coupling is a purely quantum mechanical extension of the Sagnac effect.

In this article, we take another approach to study the Sagnac effect in spin-$1/2$ particles by using the method developed in Ref. \cite{Terashima} by Terashima and Ueda, where they considered a succession of infinitesimal local Lorentz transformations and showed how the spin-$1/2$ representations of these local Lorentz transformations, i.e., the local Wigner rotations, affect the state of the spin. More specifically, we present an analogue scenario in which a spin-$1/2$ quanton \cite{Leblond}  goes through a superposition of co-rotating and counter-rotating circular paths in a rotating platform. Since there exists a difference in proper time elapsed along the two trajectories, known as the Sagnac time delay, the spin evolves to different quantum states for each path of the interferometer, what leads to the degradation of the interferometric visibility given by the generation of entanglement between the spin and momentum d.o.f., as in Ref. \cite{Zych}. So, our article follows the line of research towards exploring relativistic effects in quantum phenomena. More importantly, in contrast with the works mentioned above, which showed that a rotating space-time (or a rotating frame) cause a shift in the interference pattern, here we argue that a rotating spacetime also decreases the interferometric visibility, since the internal d.o.f gets entangled with the external d.o.f. of the quanton through the local Wigner rotations. 

It is worth emphasizing here the difference in the approach we take compared to other works in the literature. For instance, in Ref. \cite{Hassel}, the authors described the Sagnac Effect in spin-$1/2$ particles through the WKB approximation, by assuming that the waves are propagating on macroscopic paths, such that they can be treated in a semiclassical way. From these assumptions, it is possible to derive an overall phase shift induced in the wave function. A similar approach was taken by Anandan in Ref. \cite{Anandan}, where the author applied a similar formalism for the Klein-Gordon equation, and thus for a spinless particle. Therefore, one can see that the standard approach so far in the literature is to obtain a phase shift to the overall wave function, without taking into account the effects on the internal degrees of freedom of the particle. And this is exactly what the local Lorentz transformation, through the local Wigner rotation, takes into account. Beyond that, Lorentz boosts, in general, can be regarded as controlled quantum operations where momentum plays the role of the control system, while the spin is taken as the target qubit. Therefore, Lorentz boosts perform global transformations on single particle systems, since if we have a superposition of momentum states and then a Lorentz boost is performed, the spin states get entangled with the momentum states, as argued  by Peres and Terno in Ref. \cite{Peres}.

The remainder of this article is organized in the following manner. In Sec. \ref{sec:spin}, we review the spin-$1/2$ representations of the local Lorentz transformation in curved spacetimes/accelerated frames and obtain the local Wigner rotations for a quanton moving in a circular path in a rotating platform in Minkowski spacetime. In Sec. \ref{sec:intvis}, we consider a spin-$1/2$ quanton in a Sagnac interferometer and show how the interferometric visibility is affected by spacetime effects. Thereafter, in Sec. \ref{sec:con}, we present our conclusions.

\section{Methodology}
\label{sec:spin}
\subsection{Spin-$1/2$ particle states in vielbein frames}
The investigation of the dynamics of spin-$1/2$ particles in gravitational fields, or in accelerated frames, demands the use of local reference frames (LRFs). These LRFs are defined at every point of space-time using a tetrad field or vielbein, a set of four 4-vector fields which are linearly independent  \cite{Wald}. The space-time is a differential manifold $\mathcal{M}$  \cite{Carroll}. Thus, for each point $p\in\mathcal{M}$, it provides coordinate bases for the tangent, $T_{p}(\mathcal{M})$, and cotangent, $T^*_{p}(\mathcal{M})$, spaces. Theses bases are $\{\partial_{\mu}\}$ and $\{ dx^{\nu} \}$, respectively, and satisfy $dx^{\nu}(\partial_{\mu}) := \partial_{\mu}x^{\nu} = \delta^{\ \nu}_{\mu}$. So, the metric tensor can be recast as $g = g_{\mu \nu}(x) dx^{\mu} \otimes dx^{\nu}$. The elements of the metric tensor, defined by $g_{\mu \nu}(x) = g(\partial_\mu, \partial_{\nu})$, encode the gravitational field. Once the coordinate bases $\{\partial_{\mu}\} \subset T_p(\mathcal{M})$ and $\{ dx^{\nu} \} \subset T^*_p(\mathcal{M})$ are not obligatorily orthonormal, one can set up any convenient basis. For instance, one can set up an orthonormal basis with respect to the pseudo-Riemannian manifold we are working on. In view of Ref. \cite{Nakahara}, we use
\begin{align}
    & e_a = e_a^{\ \mu}(x) \partial_\mu, \ \ \ e^a = e^a_{\ \mu}(x)dx^{\mu}, \\
    & \partial_{\mu} = e^a_{\ \mu}(x)e_a, \ \ \ dx^{\mu} = e_a^{\ \mu}(x)e^a.
\end{align}  
For each $p \in \mathcal{M}$, the Minkowski metric tensor in the LRF, $\eta_{ab} = \text{diag}(-1,1,1,1)$, and the space-time metric, $g_{\mu \nu}(x)$, are connected by the tetrad field as follows:
\begin{align}
    & g_{\mu \nu}(x)e_a^{\ \mu}(x)e_b^{\ \nu}(x) = \eta_{ab},\\ &  \eta_{ab}e^a_{\ \mu}(x)e^b_{\ \nu}(x) = g_{\mu \nu}(x) \label{eq:metr},
\end{align}
with
\begin{align}
    e^a_{\ \mu}(x)e_b^{\ \mu}(x) = \delta^{a}_{\ b}, \ \ \ e^a_{\ \mu}(x)e_a^{\ \nu}(x) = \delta_{\mu}^{\ \nu}.
\end{align}

Throughout this article, Latin indices $a, b, c, d,\cdots$ are utilized to coordinates in the LRF and Greek indices $\mu, \nu, \cdots$ are used for the four general-coordinate labels. Repeated indices are summed over. The components of the vielbein transform objects from the the LRF, $x^a$, to the general coordinate system, $x^{\mu}$,  and vice versa. So, the vielbein can be utilized to move the dependence of space-time curvature of the tensor fields toward the tetrad fields. Besides, Eq. (\ref{eq:metr}) demonstrates that the vielbein embodies all the space-time curvature information encoded in the metric tensor. In addition, the vielbein $\{e_a^{\ \mu}(x), a = 0,1,2,3\}$ is a set of four 4-vector fields, and transforms under local Lorentz transformations (LLTs) in the local system. The LRF is not unique, since it continues to be local
under LLTs. Thus, a vielbein representation of a given metric is not defined in a unique manner, and different vielbein shall give the same metric tensor, provided that they are connected by LLTs \cite{Misner}.

Here, we regard the Minkowski space-time in a rotating reference frame $\Sigma$, which revolves relative to $\Sigma'$ around their common $z$-axis with constant angular velocity $\omega$. The coordinates in $\Sigma$ are related to those in $\Sigma'$ by \cite{Ryder}
\begin{align}
    t = t', \ \ r = r', \ \ z = z', \ \ \phi = \phi' - \omega t',
\end{align}
such that the invariant line element reads 
\begin{align}
    ds^2 = -(1 - \omega^2 r^2) dt^2 + dr^2 + dz^2 + r^2 d\phi^2 + 2 \omega r^2 dt d\phi. \label{eq:met}
\end{align}
We use $c = 1$. The form of the metric above is useful to define world lines of locally non-rotating observers which carry an orthonormal frame with them, i.e., we can define the following vielbein
\begin{equation}
     e^{0}_{\ t} = \ 1, \ \ e^{1}_{\ r} = 1, \ \  e^{2}_{\ z} = 1, \ \  e^{3}_{\ t} =  \omega r, \ \ \ \ e^{3}_{\ \phi} = r.
\end{equation}
The other components are all null. Hereafter, only the non-vanishing components will be shown. For the components above, the inverse elements are
\begin{equation}
    e_{0}^{\ t} = \ 1, \ \ e_{0}^{\ \phi} = - \omega, \ \ e_{1}^{\ r} = 1 \ \   e_{2}^{\ z} = 1, \ \  e_{3}^{\ \phi} =  1/r.
\end{equation}
Therefore, from the tetrad field defined above, we turn the Minkowski metric given by Eq. (\ref{eq:met}) into its usual form $ds^2 = \eta_{ab}e^a e^b$. Therefore, the observers defined by the tetrad field above are just the well known inertial frames of special relativity. One can see this by noting that, in the coordinate (rotating) frame $\Sigma$, the four velocity of such observers can be defined as $v^{\mu} := e_{0}^{\ \mu}(x) = (1, 0, 0, - \omega)$, which is rotating with respect with $\Sigma$. Besides such observers are inertial, since $A^{\mu} := v^{\nu} \nabla_{\nu} v^{\mu} = 0$, where $A^{\mu}$ is the four-acceleration of such observers and $\nabla_{\nu}$ is the covariant derivative. On the other hand, these observers are locally non-rotating since $v^a = e^a_{\ \mu} u^{\mu} = (1, 0, 0, 0)$. This type of observer is called ZAMO observer \cite{Barden} in a more general context. The introduction of such tetrad field in the Minkowski spacetime of a rotating frame may appear unnecessary or artificial. However, we want to describe spin-$1/2$ particles with respect to such observers. Thus, as mentioned above, to describe spin-$1/2$ quantons in such accelerated frames, or in gravitational fields, the introduction of a tetrad field is essential.

To define a particle with spin-$1/2$ in curved spacetimes, we have to construct the LLT such that a particle is then defined as a one-particle state furnishing the spin-$1/2$ representation of the LLT \cite{Terashima}. The construction of LLT will be revised in the next section. Now, if $p^{\mu}(x) = m u^{\mu}(x)$ represents the four-momentum of such particle, where $u^{\mu}(x)$ is the four-velocity and $p^{\mu}(x) p_{\mu}(x) = -m^2$ in the general reference frame, then the momentum in the local frame is given by $p^a(x) = e^a_{\ \mu}(x) p^{\mu}(x)$. So, the representation of a spin $\sigma$ quantum state, with momentum $p^a(x)$ as observed from the position $x^a = e^a_{\ \mu}(x) x^{\mu}$ of the LRF defined by $ e^a_{\ \mu}(x)$ in the spacetime $\mathcal{M}$ with metric $g_{\mu \nu}(x)$, is \cite{Lanzagorta}:
\begin{align}
 \ket{p^a(x), \sigma; x} := \ket{p^a(x), \sigma; x^{a}, e^a_{\ \mu}(x), g_{\mu \nu}(x)}.
\end{align}

It's worth pointing out that the description of a spin-$1/2$ particle quantum state can only be given in relation to the vielbein and the local structure it describes, since $e_0^{\ \mu}(x)$ is a time-like vector field defined in each point of the space-time and produces a global time coordinate, thus making the space-time time orientable \cite{Wald}. By definition, the state $\ket{p^a(x), \sigma; x}$ transforms as the spin-$1/2$ representation under the LLT. In Special Relativity, it is known that the spin-$1/2$ particle state $\ket{p^a, \sigma}$ transforms under a component of the Poincar\'e group $\Lambda^{a}_{\ b}$ as a unitary representation given as follows \cite{Weinberg}:
\begin{equation}
    U(\Lambda)\ket{p^a, \sigma} = \sum_{\lambda} D_{\lambda \sigma} (W(\Lambda,p)) \ket{\Lambda p^a, \lambda}.
\end{equation}
Above $D_{\lambda, \sigma}(W(\Lambda,p))$ is a unitary representation of Wigner's little group. The elements of this group are the well known Wigner's rotations (WR) $W^{a}_{\ b} (\Lambda,p)$ \cite{Eugene}. The subscripts can be suppressed and sometimes we will write $U(\Lambda) \ket{p^a, \sigma} = \ket{\Lambda p^a} \otimes D (W(\Lambda, p)) \ket{\sigma}.$ Thus, one can see that the particle's spin is rotated and it is controlled by the momentum of the particle. In contrast, in general relativity a one-particle state forms a local representation of the inhomogeneous Lorentz group at each point $p \in \mathcal{M}$, i.e.,
\begin{equation}
    U(\Lambda(x))\ket{p^a(x), \sigma;x} = \sum_{\lambda} D_{\lambda \sigma}(W(x)) \ket{\Lambda p^a(x), \lambda;x} \label{eq:unit}.
\end{equation}
In this last equation, $W(x) := W(\Lambda(x), p(x))$ is a local WR.

\subsection{Local Lorentz transformations and local Wigner rotations}
In this section, we recapitulate the construction of LLTs and its spin-$1/2$ representation and obtain such quantities for a quanton moving in a circular path in a rotating platform in Minkowski spacetime. Following Ref. \cite{Terashima}, in the local tetrad frame at point $p$ with coordinates $x^a = e^a_{\ \mu}(x) x^{\mu}$, the momentum of the particle is given by $p^a(x) = e^a_{\ \mu}(x) p^{\mu}(x)$. Passed an infinitesimal interval of proper time $d \tau$, the particle moves to the new point $x'^{\mu} = x^{\mu} + u^{\mu} d\tau$. In this way, the momentum in the LRF at the new point is given by $p^a(x') = p^a(x) + \delta p^a(x)$. Such infinitesimal change can be decomposed as
\begin{equation}
    \delta p^a(x) = e^a_{\ \mu}(x) \delta p^{\mu}(x) + \delta e^a_{\ \mu}(x)p^{\mu}(x). \label{eq:momen}
\end{equation}
The variation $\delta p^{\mu}(x)$ can be attributed to an external non-gravitational force 
\begin{align}
     \delta p^{\mu}(x)&  = u^{\nu}(x) \nabla_{\nu} p^{\mu}(x) d\tau = m a^{\mu}(x) d\tau  \\
     & =  - \frac{1}{m}(a^{\mu}(x)p_{\nu}(x) - p^{\mu}(x)a_{\nu}(x))p^{\nu}(x) d\tau, \nonumber 
\end{align}
where it was used the normalization condition for $p^{\mu}(x)$ and the fact that $p^{\mu}(x)a_{\mu}(x) = 0$. As such, the variation of the vielbein is due to space-time geometrical effects and it is obtained as follows
\begin{align}
    \delta e^a_{\ \mu}(x) & = u^{\nu}(x) \nabla_{\nu}e^a_{\ \mu}(x) d\tau \nonumber \\
    & = - u^{\nu}(x) \omega_{\nu \ b}^{\ a}(x) e^b_{\ \mu}(x)d \tau.
\end{align}
In this equation, $\omega_{\nu \ b}^{\ a} := e^{a}_{\ \lambda} \nabla_{\nu} e_{b}^{\ \lambda} = - e_{b}^{\ \lambda} \nabla_{\nu} e^{a}_{\ \lambda} $ is the spin connection \cite{Chandra}. Putting together these results and substituting in Eq. (\ref{eq:momen}), one obtains
\begin{equation}
    \delta p^a(x) = \lambda^{a}_{\ b}(x)p^{b}(x) d\tau \label{eq:momvar}, 
\end{equation}
where
\begin{align}
    \lambda^{a}_{\ b}(x) & = - (a^{a}(x)u_{b}(x) - u^{a}(x)a_{b}(x)) + \chi^{a}_{\ b} \label{eq:infloc}
\end{align}
with $\chi^{a}_{\ b} :=  - u^{\nu}(x) \omega_{\nu \ b}^{\ a}(x)$.  Eqs. (\ref{eq:momvar}) and (\ref{eq:infloc}) constitute an infinitesimal LLT since,  as the particle moves along its world line during an infinitesimal proper time interval $d \tau$, the momentum in the local frame transforms as $p^{a}(x) = \Lambda^{a}_{\ b}(x) p^b(x)$ where $\Lambda^{a}_{\ b}(x) = \delta^{a}_{\ b} + \lambda^{a}_{\ b}(x)d \tau$ \cite{Lanzagorta}. 

In our case, the quanton follows a circular path according to the reference frames defined by the vielbein with the local four-velocity given by
\begin{align}
    u^a = (\cosh \xi, 0, 0, \sinh \xi),
\end{align}
such that $u = dx^3/dx^0 = u^3/u^0 = \tanh \xi$. In the coordinate frame $\Sigma$, the four-velocity of the quanton is given by
\begin{align}
    u^{\mu} = e_a^{\ \mu}(x) u^a = (\cosh \xi, 0, 0, \frac{\sinh \xi}{r} - \omega \cosh \xi),
\end{align}
such that $\Omega = d \phi / d t = \frac{u}{r} - \omega$. It's worth to emphasize that the quanton is moving in a circular path in the rotating frame $\Sigma$ as well. Besides, if we set the velocity of the particle in the local frames as $u = r\omega$, the quanton stays still with respect to the coordinate frame $\Sigma$, since it's rotating with the same angular velocity $\omega$. 

The circular path followed by the quanton with four-velocity $u^{\mu}$ is not a geodesic and, therefore, it's necessary a force to maintain the quanton in such circular path with such four-velocity. The non-zero component of the acceleration due to the external force is given by:
\begin{align}
    a^r = u^{\nu} \nabla_{\nu} u^r = - \frac{\sinh^2 \xi}{r}.
\end{align}
Besides, one can show that the only non-zero infinitesimal LLT is given by
\begin{align}
    \lambda^{1}_{\ 3} = - \lambda^{3}_{\ 1} = \frac{\sinh \xi \cosh^2 \xi}{r},
\end{align}
where $\chi^{1}_{\ 3} = \sinh \xi/r$.

\begin{figure}[t]
\centering
\includegraphics[scale=0.45]{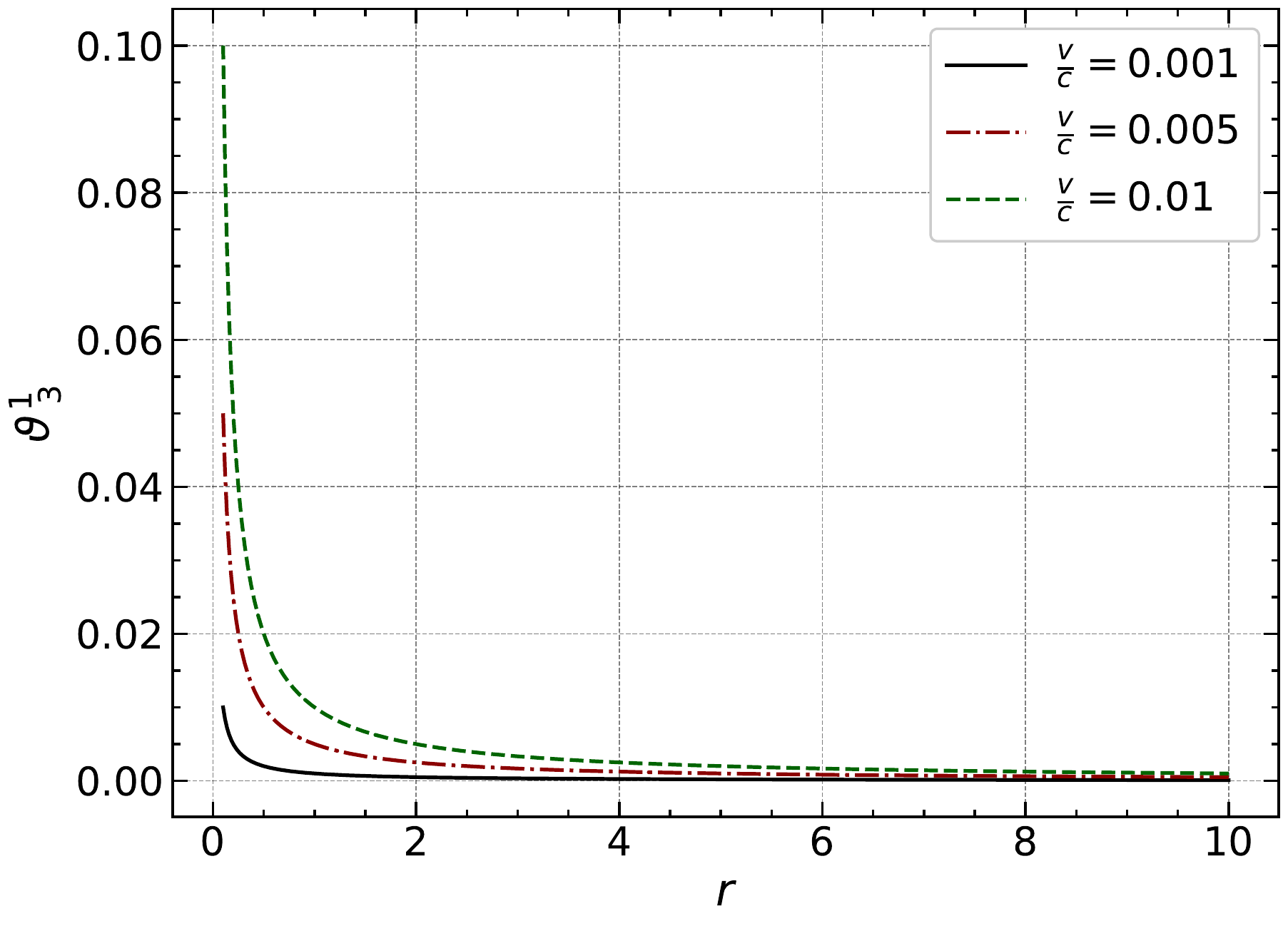}
\caption{The local Wigner rotation as a function of the radial coordinate, $r$.}
\label{fig:thetami}
\end{figure}

Now, as mentioned before, by using a unitary representation of the LLT, in the point $x^{\mu}$, the state of the quanton is described by $\ket{p^a(x), \sigma; x}$ such that, in the LRF at the point $x'^{\mu}$, the state of the quanton is now described by $U(\Lambda(x)) \ket{p^a(x), \sigma; x}$ defined by Eq. (\ref{eq:unit}). Thus, the state of the spin changes in a local manner as the quanton moves from $x^{\mu} \to x'^{\mu}$. For the infinitesimal LLT, the infinitesimal local WR reads: \cite{Kilian} $ W^{a}_{\ b}(x) = \delta^{a}_{\ b} + \vartheta^{a}_{\ b} d \tau,$ where 
\begin{equation}
    \vartheta^{i}_{\ j}(x) = \lambda^{i}_{\ j}(x) + \frac{\lambda^{i}_{\ 0}(x)p_j(x) - \lambda_{j0}(x)p^i(x)}{p^0(x) + m},
\end{equation}
whereas all other terms vanish. In our case, a straightforward calculation shows that the only non-vanishing local WR is given by
\begin{align}
    \vartheta^{1}_{\ 3} = - \vartheta^{3}_{\ 1} = \frac{\sinh \xi \cosh \xi}{r},
\end{align}
which corresponds to a rotation over the z-axis. It is worth noticing that $\chi^{1}_{\ 3} \neq \lambda^{1}_{\ 3} \neq \vartheta^{1}_{\ 3}$, where these two non-equalities result from the acceleration of the quanton and the boost part of $\lambda^a_{\ b}$, respectively \cite{Terashima}. In Fig. \ref{fig:thetami}, we plotted the local Wigner rotation as a function of $r$ for different values of $v/c$. Finally, the difference $\vartheta^3_{\ 1} - \chi^3_{\ 1}$ gives rise to the Thomas precession. For instance, when $v << c$, the precession angle per $dt = \cosh \xi d \tau$ becomes $(\vartheta^3_{\ 1} - \chi^3_{\ 1})d \tau \approx - \frac{u a}{2}dt$, where $a := \abs{a^r}$ \cite{Terashima}.

Hence, the two-spinor representation of the infinitesimal Wigner rotation is then given by
\begin{align}
    D(W(x)) & = I_{2 \times 2} + \frac{i}{4} \sum_{i,j,k = 1}^{3} \epsilon_{ijk} \vartheta_{ij}(x) \sigma_k d \tau \nonumber \\
    & = I_{2 \times 2} + \frac{i}{2} \boldsymbol{\vartheta} \cdot \boldsymbol{\sigma} d\tau, \label{eq:wigner}
\end{align}
where $I_{2 \times 2}$ stands for the identity matrix and $\{\sigma_k\}_{k = 1}^3$ are the well known Pauli matrices. Furthermore, the WR for a particle moving over a finite proper time interval is obtained by iterating the infinitesimal WR \cite{Terashima}, and the spin-$1/2$ representation for a given finite proper time is obtained by iterating Eq. (\ref{eq:wigner}):
\begin{equation}
    D(W(x, \tau)) = \mathcal{T}e^{\frac{i}{2}\int_0^{\tau} \boldsymbol{\vartheta} \cdot \boldsymbol{\sigma} d\tau'}. \label{eq:time}
\end{equation}
Above $\mathcal{T}$ is the time-ordering operator \cite{Terashima}.

\section{Sagnac Effect for spin-$1/2$ particles} 
\label{sec:intvis}
In this section, we shall investigate the Sagnac effect for spin-$1/2$ particles through local WRs and we will show that there exists a degradation of the interferometric visibility given by the generation of entanglement between the spin d.o.f. and the momenta under local WRs as a spin-$1/2$ quanton goes through a superposition of co-rotating and counter-rotating circular paths in a rotating platform. To make our investigation easier, we begin by assuming that momenta can be treated as discrete variables, as in Refs. \cite{Zych, Terashima}. As pointed out in Ref. \cite{Nasr}, this can be justified once that it's possible to assume that the quanton wave-packet has a mean centroid that one can regard to describe the motion of the center of mass in each superposed path such that the quanton's momenta is spread duly around its mean value, given that the mean value moves along a given path $x^{\mu}(\tau)$ in the Minkowski spacetime. Therefore, the four-momentum of the mean value as measured in a local frame will be $p^a(x) = e^a_{\ \mu}(x)p^{\mu}(x)$. Hence, the local observer deals only with the mean values. First, the mean value of the momenta which is directly related to the four-velocity of the corresponding circular geodesic. Therefore, both distributions are distinguishable (do not overlap) and centered around different momentum values such that it is possible to represent them by orthogonal vectors. Second, the mean value of the center of mass in the position basis which is also a Gaussian distribution with the mean value corresponding to the coordinates $x^a$ of the circular geodesic.

\begin{figure}[t]
\centering
\includegraphics[scale=0.6]{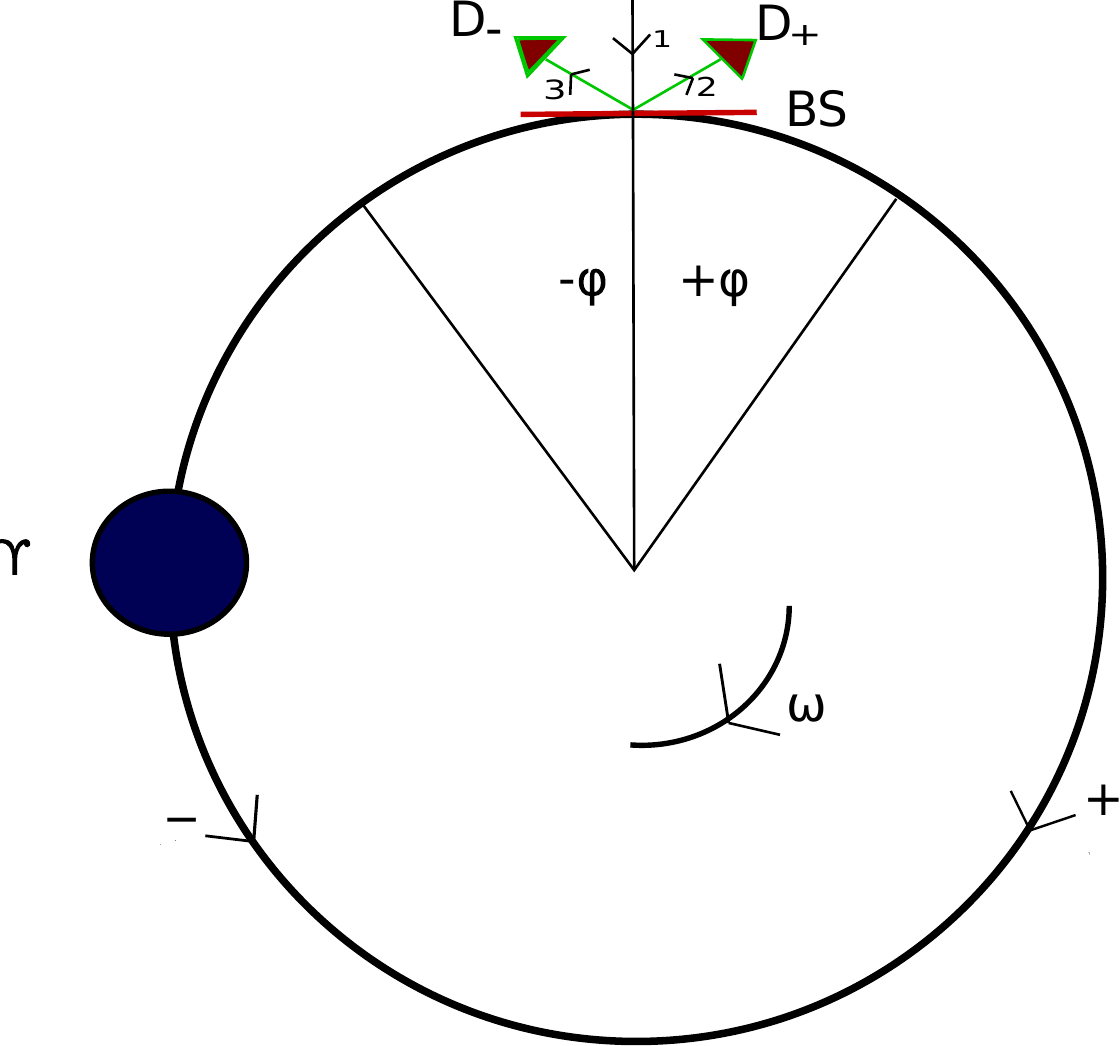}
\caption{A spin-$1/2$ quanton in coherent superposition of circular paths in a rotating platform in Minkowski spacetime. The apparatus consists of a beam splitter $BS$, a phase shifter, which gives a controllable phase $\Upsilon$, and detectors $D_{\pm}$ in a rotating platform.}
\label{fig:sagnac}
\end{figure}

Fig. \ref{fig:sagnac} represents a spin-$1/2$ quanton in coherent superposition of circular paths in a rotating platform. The apparatus consists of a beam splitter $BS$, a phase shifter, which gives a controllable phase $\Upsilon$ that can be the angular velocity $\omega$ of the platform, and two detectors $D_{\pm}$. The initial state of the quanton, before the $BS$, is given by $  \ket{\Psi_{i}} = \ket{p_i} \otimes \ket{\tau_i } = \frac{1}{\sqrt{2}}\ket{p_i} \otimes (\ket{\uparrow} + \ket{\downarrow})$, with the 1-axis being local quantization of the spin. Just after the $BS$, the state changes to
\begin{align}
    \ket{\Psi} & = \frac{1}{2}(\ket{p_{+};0} + i\ket{p_{-};0})  \otimes  (\ket{\uparrow} + \ket{\downarrow}) \label{eq:state2},
\end{align}
where $\phi = 0$ corresponds to the coordinate of the point where the quanton was putted in a coherent superposition in opposite directions with constant four-velocity $u^a_{\pm} = (e^{0}_{\ \mu} u^{\mu}_{\pm}, 0, 0, e^{3}_{\ \mu} u^{\mu}_{\pm})$.  Here, we assumed that the beam-splitter $BS$ do not affect the spin degree of freedom. However, it is possible to consider the beam-splitter as a Stern-Gerlach apparatus, such that the spin d.o.f is coupled to the momentum d.o.f. 

After some interval of proper time $d\tau$, the quanton has travel along its circular paths such that the spin-$1/2$ representation of the finite WR is given by 
\begin{align}
    D(W(\pm,\tau)) = e^{- \frac{i}{2}\sigma_2 \vartheta^1_{\ 3} \int_{\pm} d \tau},  \label{eq:wigrot}
\end{align}
where $+$ refers to the co-rotating circular path while $-$ refers to the counter-rotating circular path. The time-ordering operator is not needed, once $\vartheta^{1}_{\ 3}$ is constant along the circular paths. Therefore, the state of the quanton in the local frame at point $\phi = 2\pi$, right before $BS$, is given 
\begin{align}
U(\Lambda)\ket{\Psi}  =& \frac{1}{2} \ket{p_{+};2 \pi} \otimes e^{- \frac{i}{2}\sigma_y  \vartheta^1_{\ 3} \int_{+} d \tau}(\ket{\uparrow} + \ket{\downarrow})  \nonumber \\
& + \frac{i e^{i \Upsilon}}{2}\ket{p_{-};2\pi}\otimes e^{-\frac{i}{2}\sigma_y   \vartheta^1_{\ 3} \int_{-} d \tau}(\ket{\uparrow} + \ket{\downarrow}),   \label{eq:state1}
\end{align}
which, in general, is an entangled state. The detection probabilities corresponding to Eq. (\ref{eq:state1}), after the $BS$, are given by
\begin{align}
    P_{\pm} = \frac{1}{2}\Big(1 \mp \cos \Big(\frac{\vartheta^1_{\ 3}\Delta\tau}{2} \Big) \cos \Upsilon \Big),
\end{align}
where $\Delta \tau := \int_+ d \tau -  \int_- d \tau$ is the difference of the proper time between the two circular paths.

By varying the controllable phase shift $\Upsilon$, the probabilities $P_{\pm}$ are made to oscillate with amplitude $\mathcal{V}$, which is called the interferometric visibility and can be calculated using \cite{Zych}:
\begin{align}
    \mathcal{V} & = \abs{\bra{\tau_i} e^{-\frac{i}{2}\sigma_y \vartheta^1_{\ 3}\Delta\tau} \ket{\tau_i}} \label{eq:visibi} \\
    & = \abs{ \cos \Big(\frac{\vartheta^1_{\ 3}\Delta\tau}{2} \Big)}, \label{eq:cosi}
\end{align}
where $\ket{\tau_i} = \frac{1}{\sqrt{2}}(\ket{\uparrow} + \ket{\downarrow})$ is the initial state of the clock. In addition, the difference $\Delta \tau$ is the Sagnac time delay that for matter beams can be calculated through \cite{Zendri, Tartag}
\begin{align}
    \Delta \tau & = - 2 \sqrt{- g_{tt}(x_{det})} \oint \frac{g_{t \phi}}{g_{tt}} d\phi = \frac{4 \pi r^2 \omega}{\sqrt{1 - \omega^2 r^2}} \approx 4 A \omega,
\end{align}
where $g_{tt}(x_{det})$ is the $tt$-component of the metric evaluated at detector's position, which is also at a distance $r$ from the center of the disk and $A$ is the area embraced by the arms of the interferometer. Besides, it is important to say that the total angle $\Theta_{\pm}:= \vartheta^1_{\ 3} \int_{\pm} d \tau$ reflects the "trivial rotation" of $2 \pi$, that the spin undergoes as it completes the circular orbit, and the rotation due to spacetime effects \cite{Terashima}. Therefore, it's common to compensate the trivial rotation angle of $2 \pi$ to obtain the total Wigner rotation solely by spacetime effects by defining $\Omega_{\pm} = \Theta_{\pm} - 2\pi$. However, in our case, the trivial rotation of each circular does not affect the visibility, since Eq. (\ref{eq:cosi}) will be the same for any argument multiple of $2\pi$, i.e., $\vartheta^1_{\ 3}\Delta\tau \pm 2n \pi$, where $n$ is an integer.

Now, without the entanglement between the spin and the momentum, the visibility reaches its maximum possible value, i.e., $\mathcal{V} = 1$. However, with the introduction of the internal d.o.f. and its entanglement with the momentum results in a change in the interferometric visibility. In this case, the difference of the total WR for each branch of the interferometer can be assigned to the Sagnac time delay, since $\vartheta^1_{\ 3}$ is the same for both paths. This is the main result of this work, i.e., the fact that, for spin-$1/2$ particles, the Sagnac time delay also affects the interferometric visibility according to the framework developed in Ref. \cite{Terashima} through local WR, since the spin evolves to different quantum states for each path of the interferometer, as one can see in Eq. (\ref{eq:state1}), which leads to the degradation of the interferometric visibility given by the generation of entanglement between the spin d.o.f. and the momenta.

\begin{figure}[t]
    \centering
    \subfigure[$\mathcal{V}$ as a function of $\omega$, for the parameters $r = 3m$ and $\frac{v}{c} = 0.6 \times 10^{-5}$.]{{\includegraphics[scale = 0.4]{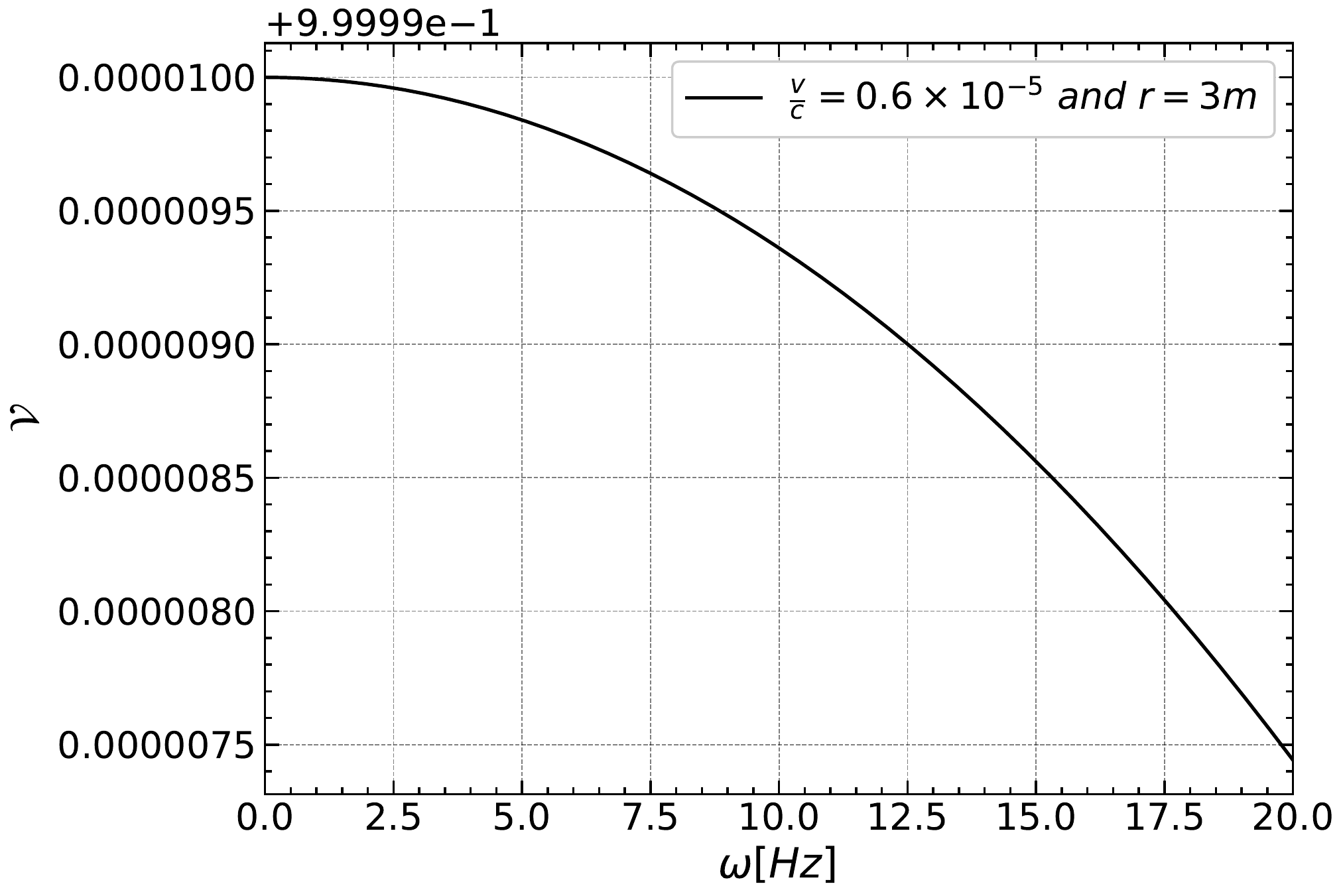}{\label{fig:e}} }}
    \subfigure[$S_{vn}(\rho_{\Lambda p}) = S_{vn}(\rho_{\Lambda s})$ as a function of $\omega$, for the parameters $r = 3m$ and $\frac{v}{c} = 0.6 \times 10^{-5}$.]{{\includegraphics[scale = 0.4]{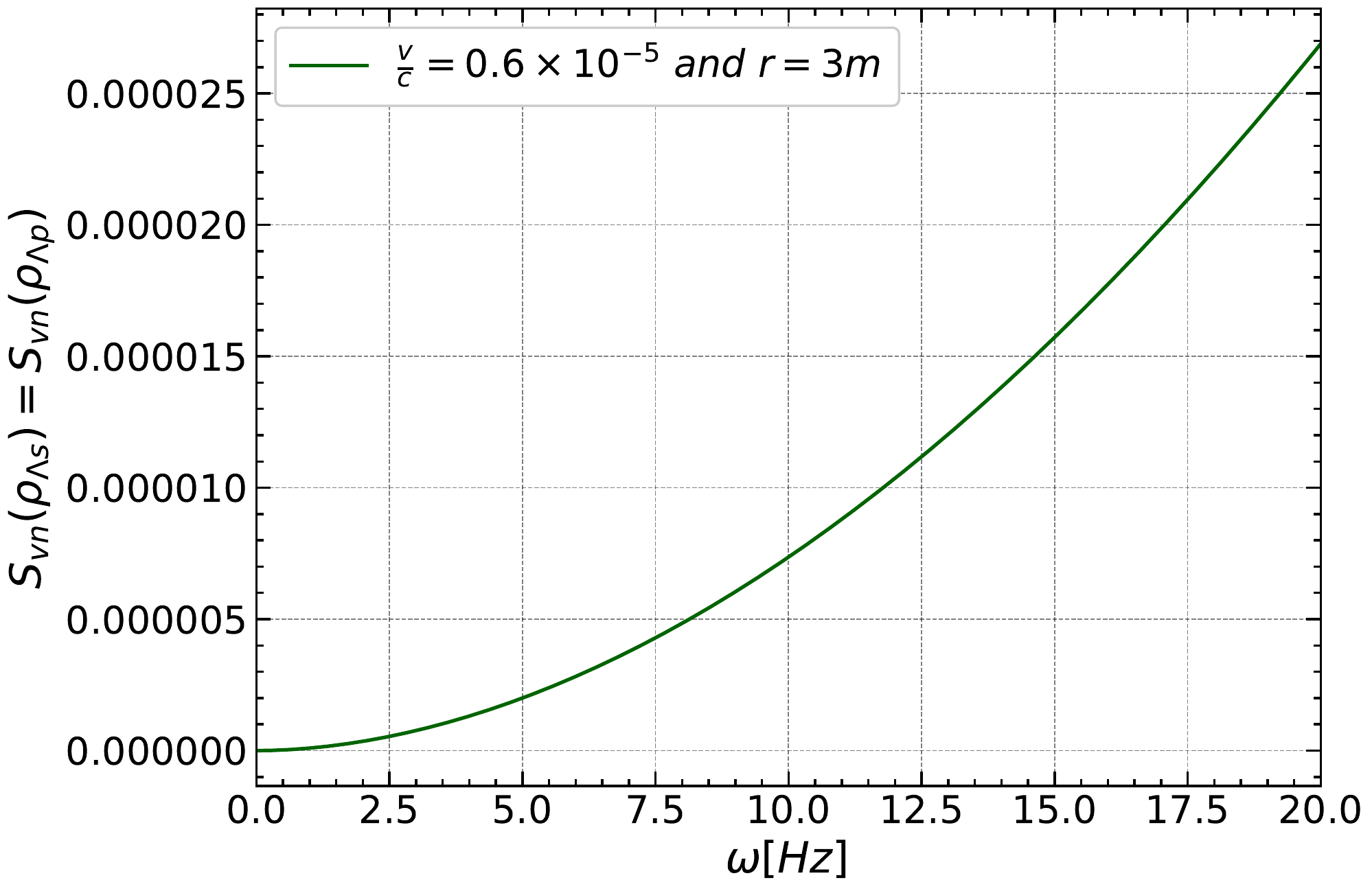}{\label{fig:a}} }}

  \caption{Interferometric visibility, $\mathcal{V}$, and entanglement entropy, $S_{vn}(\rho_{\Lambda p}) = S_{vn}(\rho_{\Lambda s})$, as a function $\omega$.}    
\end{figure}

\begin{figure}[t]
    \centering
    \subfigure[$\mathcal{V}$ as a function of $r$, for the parameters $\omega = 10 Hz$ and $\frac{v}{c} = 0.6 \times 10^{-5}$.]{{\includegraphics[scale = 0.4]{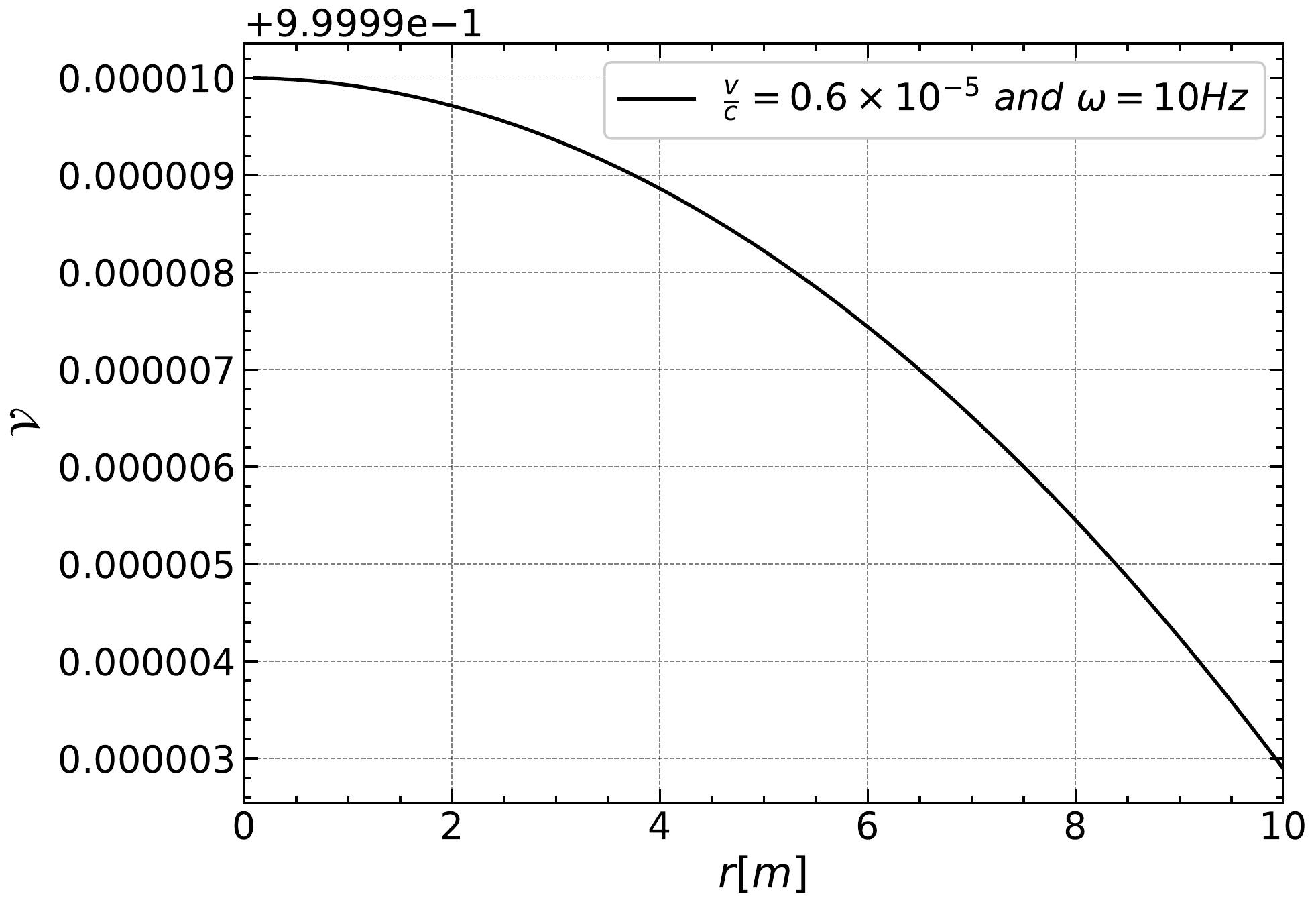}{\label{fig:f}} }}
    \subfigure[$S_{vn}(\rho_{\Lambda p}) = S_{vn}(\rho_{\Lambda s})$ as a function of $r$, for the parameters $\omega = 10 Hz$ and $\frac{v}{c} = 0.6 \times 10^{-5}$.]{{\includegraphics[scale = 0.4]{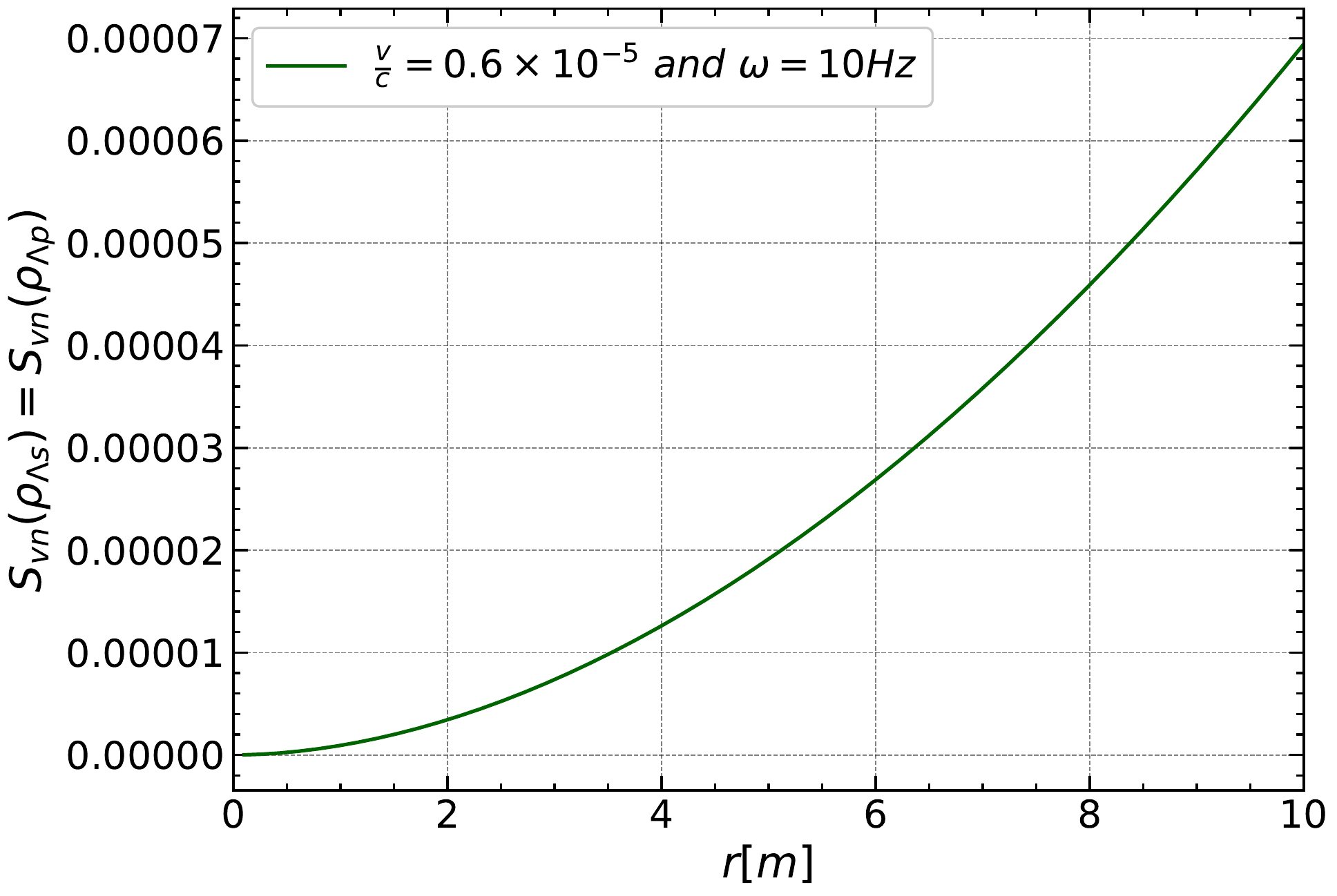}{\label{fig:b}} }}

\caption{Interferometric visibility, $\mathcal{V}$, and entanglement entropy, $S_{vn}(\rho_{\Lambda p}) = S_{vn}(\rho_{\Lambda s})$, as a function $r$.}    
\end{figure}

Finally, in Fig. \ref{fig:e} we plotted the visibility as a function of $\omega$ for the parameters $r = 3m$ and $\frac{v}{c} = 0.6 \times 10^{-5}$, where the velocity of the particle corresponds to typical velocities of thermal neutrons \cite{Sam}, which is of order $v \approx 2000 m/s$, as well the values of $\omega$ and $r$ are in the same order of those tested for photons in Ref. \cite{Toros}. As well, in Fig. \ref{fig:f}, it is shown $\mathcal{V}$ as a function of $r$, for the parameters $\omega = 10 Hz$ and $\frac{v}{c} = 0.6 \times 10^{-5}$. While, in Figs. \ref{fig:a} and \ref{fig:b} we plotted the von Neumann entropy of the reduced density matrix of the state given by Eq. (\ref{eq:state1}), i.e.,  $S_{vn}(\rho_{\Lambda s}) = S_{vn}(\rho_{\Lambda p})$, as a function of $\omega$ and $r$, respectively, where $\rho_{\Lambda s} = \Tr_p(U(\Lambda) \ketbra{\Psi} U^{\dagger}(\Lambda))$ denotes the reduced density matrix of the spin and similar for the momenta. Since the bipartite quantum system is in a pure state given by Eq. (\ref{eq:state1}), the von Neumann entropy of the reduced states is a well known entanglement monotone \cite{Hor}. Thus, one can see that the decrease of the interferometric visibility due to the entanglement between the external and internal d.o.f are very tiny and difficult to be measured. Besides, it is worthwhile to point out that, as in Ref. \cite{Costa}, a `clock' with a Hilbert space of finite dimension has a periodic time evolution. So, it is to be anticipated that the visibility oscillates periodically as a function of the difference of the proper times elapsed in the two arms of the interferometer. This will also happen here if we are able to control spin-$1/2$ particles in a Sagnac interferometer with velocities bigger than the velocity of typical thermal neutrons, or to increase any of the other two parameters: $r$ and $\omega$. Finally, it is worth mentioning that, if the beam splitter entangles the momentum and spin, then the visibility after $BS$, in principle, is zero because the which-path information is encoded in the spin d.o.f Therefore, the action of $U(\Lambda)$ in this case will be to degrade the entanglement between both degrees of freedom, and to increase the quantum coherence of both degrees of freedom as the quanton is travelling in the arms of the interferometer, since, in this scenario, entanglement and coherence are complementary quantities, as already shown by us in Ref. \cite{Jonas}.

\section{Conclusions}
\label{sec:con}
In this article, we studied the Sagnac effect for spin-$1/2$ particles through the local Wigner rotations according to the framework developed in Ref.  \cite{Terashima}. The main result of our work is the realization that the dissimilarity in the proper time elapsed along the two paths, known as the Sagnac time delay, makes the spin evolve to different states for each branch of the two-arms interferometer. This leads to the degradation of the interferometric visibility given by the creation of entanglement between the spin d.o.f. and the momenta, as argued in Ref. \cite{Zych}. However, the decrease of the interferometric visibility due to the entanglement between the external and internal d.o.f. are very tiny for velocities typical of thermal neutrons. Finally, it is worth mentioning that it is possible to explore the Sagnac effect for spin-$1/2$ particles in another spacetimes, as those regarded in Refs. \cite{Tartagli, Luca}.

\begin{acknowledgments}
This work was supported by the Coordena\c{c}\~ao de Aperfei\c{c}oamento de Pessoal de N\'ivel Superior (CAPES), process 88882.427924/2019-01, and by the Instituto Nacional de Ci\^encia e Tecnologia de Informa\c{c}\~ao Qu\^antica (INCT-IQ), process 465469/2014-0.
\end{acknowledgments}


\end{document}